\documentclass[journal=jpccck,manuscript=article]{achemso}

\usepackage{graphicx,epsfig}

\author{Ersen Mete}
\email{emete@balikesir.edu.tr}
\phone{+90 (0)266 6121000}
\fax{+90 (0)266 6121215}
\affiliation{Deparment of Physics, Bal{\i}kesir University, Bal{\i}kesir 10145, Turkey}

\author{\.{I}lker Demiro\u{g}lu}
\affiliation{Department of Chemistry, Middle East Technical University, Ankara 06531, Turkey}

\author{M. Fatih Dan{\i}\c{s}man}
\affiliation{Department of Chemistry, Middle East Technical University, Ankara 06531, Turkey}

\author{\c{S}inasi Ellialt{\i}o\u{g}lu}
\affiliation{Department of Physics, Middle East Technical University, Ankara 06531, Turkey}

\title{Pentacene multilayers on Ag(111) surface}

\begin{document}

\begin{abstract}
The structural profiles and electronic properties of pentacene
(C$_{22}$H$_{14}$) multilayers on Ag(111) surface has been studied
within the density functional theory (DFT) framework. We have performed
first-principle total energy calculations based on the projector augmented
wave (PAW) method to investigate the initial growth patterns of pentacene
(Pn) on Ag(111) surface. In its bulk phase, pentacene crystallizes with
a triclinic symmetry while a thin film phase having an orthorhombic unit
cell is energetically less favorable by 0.12 eV/cell. Pentacene prefers
to stay planar on Ag(111) surface and aligns perfectly along silver rows
without any molecular deformation at a height of 3.9 {\AA}. At one monolayer
(ML) coverage the separation between the molecular layer and the surface
plane extends to 4.1 {\AA} due to intermolecular interactions weakening
surface--pentacene attraction. While the first ML remains flat, the molecules
on a second full pentacene layer deposited on the surface rearrange so that
they become skewed with respect to each other. This adsorption mode is
energetically more preferable than the one for which the molecules form a
flat pentacene layer by an energy difference similar to that obtained for bulk
and thin film phases. Moreover, as new layers added, pentacenes assemble
to maintain this tilting for 3 and 4 ML similar to its bulk phase while the
contact layer always remains planar. Therefore, our calculations indicate
bulk-like initial stages for the growth pattern.
\end{abstract}

%\pacs{85.65.+h, 73.20.Hb, 68.43.Bc}
\newpage

\section{Introduction}

Due to its use in thin film transistor (TFT) applications pentacene is continuing
to enjoy being the subject of extensive research. Since most TFT's employ
SiO$_2$ as the dielectric layer, initial studies on pentacene focused mainly
on gaining a thorough understanding of structural~\cite{Dimitrakopoulos,Ruiz1,Ruiz2,Mattheus1}
and electronic~\cite{Coropceanu,Fukagawa,Anthony1,Troisi} properties of
thin films of this molecule on SiO$_2$ surfaces and in turn achieving the
best device performance by optimizing~\cite{Yanagisawa,DeAngelis,Locklin}
these properties. As a result of this heavy research effort in the last
decade, fabricating pentacene TFT's with hole mobilities of more than
1 cm$^2$/Vs  have become an almost routine process~\cite{Kelley}. Nevertheless
there are still, fundamental issues to be resolved such as the dependence
of the charge mobility on the film thickness~\cite{Ruiz3} and areas open
to improvement like modification of the substrate surfaces with buffer
layers~\cite{Anthony1} or the pentacene itself with functional
groups~\cite{Anthony1,Anthony2}.

Another very critical but relatively less well understood subject is the
growth mechanism of pentacene films on metal substrates. Understanding the
pentacene film growth on gold and silver surfaces is particularly important
since these metals constitute the electrode material in most TFT's and the
device performance is directly related to the charge transfer efficiency
between the electrodes and the organic film. Though both
experimental~\cite{Kang1,Kang2,France1,France2,Beernink,Kafer1,Kafer2,
Casalis,Danisman,Eremtchenko,Zhang,Dougherty,Floreano,Zheng,Wang} and
theoretical~\cite{Wang,Lee1,Lee2} research in this field has been recently
intensified, there are contradictory results in the literature and the
growth modes of pentacene thin films on Au(111) and Ag(111) surfaces are
continuing to be a matter of debate.  This is mostly due to the relatively
strong (when compared with SiO$_2$) interaction of the metal surfaces with
the pentacene molecule. As a result of this strong interaction pentacene
adopts many different monolayer and multilayer phases on metal surfaces
which are energetically and structurally very close to each other.
For example on Au(111) several different low density monolayer phases
and an identical full coverage phase have been reported by different
groups~\cite{Kang1,Kang2,France1,France2,Kafer1}. However, in case of
the multilayers while Kang et al.~\cite{Kang1,Kang2} report a layer by
layer growth of lying down pentacene molecules, Beernink et al.~\cite{Beernink}
report strong dewetting starting from the second layer and growth of bulk
like pentacene crystals. For pentacene films on Ag(111) surfaces, while
Eremtchenko et al.~\cite{Eremtchenko} and Dougherty et al.~\cite{Dougherty}
report a bilayer film formation, where an ordered (second) layer (which
follows the symmetry of the Ag(111) surface) forms on top of a disordered
(2D gas phase) first layer at room temperature, K\"{a}fer et al.~\cite{Kafer2}
report the formation of bulk-like pentacene structures immediately after
the first monolayer. So on both surfaces the growth mechanism of pentacene
films is still not completely clear. In addition, if the above mentioned 2D
gas phase mechanism for Ag(111) is really correct then questions like ``Why
does pentacene behave completely differently on seemingly similar surfaces,
Ag(111) and Au(111)?'' and ``How does the symmetry of the substrate affect the
bilayer film structure?'' arise.

In spite of this richness of experimental studies and points in need of
clarification, pentacene films on Ag(111) or Au(111) surfaces have not,
yet, been studied theoretically. Theoretical work regarding pentacene
films were mostly performed on other metal surfaces, such as
Cu(001)~\cite{Satta}, Cu(110)~\cite{Ample}, Ag(110)~\cite{Wang}, at
semi-empiric level and Al(100)~\cite{Simeoni}, Cu(100)~\cite{Ferretti},
Cu(119)~\cite{Baldacchini}, Fe(100)~\cite{Sun},  Au(001)~\cite{Lee1,Lee2}
at DFT level. These studies in general addressed two points concerning the
first layer of pentacene film/molecule: (1) Determination of the most stable
adsorption site/geometry, and (2) determination of the strength of electronic
interaction/coupling between the substrate and the molecule. In these studies
either the most stable configuration was found to be pentacene lying flat on
the surface~\cite{Wang} or the calculations were started with this assumption.
In terms of the electronic interactions, the DFT studies performed using GGA
functionals found considerable aromatic-$\pi$-system metal substrate
interaction~\cite{Ferretti,Baldacchini} on Cu surfaces, hinting at
chemisorption. On Au(111)~\cite{Lee2} and Al(100)~\cite{Simeoni} however,
while LDA functionals resulted in strong interactions, in the form of
broadening and splitting of  $\pi$-molecular orbitals, GGA functionals are
reported to result in much weaker interactions, in accord with a physisorption
mechanism. Theoretical studies concerning the further stages of pentacene film
growth on metal surfaces, however, like second layer structure/energetics or
the thin film crystal/electronic structure, are very few and at semi-empiric
level~\cite{Satta}. Instead, theoretical works regarding pentacene films are
primarily focused on the electronic structure of different pentacene
polymorphs observed mainly on SiO$_2$ surfaces, one being the famous
``thin film phase''~\cite{Troisi,Parisse,Doi,Nabok,Mattheus2}.

Hence, a theoretical study of growth mechanism and electronic properties of
pentacene films on Ag(111) and Au(111) may, (1) help resolve the experimental
contradictions mentioned above and (2) fill a gap in the theoretical
literature and enable a comparison of these systems with other pentacene
films. As a first attempt to this end, here we present the results of our
work on the structural and electronic properties of monolayer and multilayer
films of pentacene on Ag(111) at DFT level. First we discuss full coverage
monolayer film in the light of experimental results. We compare the
adsorption geometries and the corresponding density of states we found with
the experimental results reported so far. Then we present our results
regarding two and three monolayers of pentacene film and discuss how the
crystal and electronic structure of the Ag(111)/pentacene interface and the
film evolves with coverage. We conclude with an overall summary and
discussion of the results.

\section{Method}

We performed total energy density functional theory (DFT) calculations
using the projector-augmented wave (PAW) method~\cite{paw1,paw2} within
the generalized gradient approximation (GGA) by employing the
Perdew--Burke--Ernzerhoff (PBE)~\cite{pbe} exchange--correlation energy
functional as implemented in the Vienna Ab Initio Simulation Package
(VASP)~\cite{vasp}.

For consistency, we used a kinetic energy cutoff of 370 eV for the plane
wave expansion of single particle wavefunctions in all the calculations.
Electronic ground states has been determined by requiring a total-energy
convergence up to a tolerance value smaller than 0.1 meV. We used a
conjugate-gradient algorithm, in all geometry optimization calculations,
based on the reduction in the Hellman--Feynman forces on each constituent
atom to less than 10 meV/{\AA}.

We examined two previously known polymorphs of Pn lattice: the
bulk~\cite{Campbell} and the thin film~\cite{Parisse} phases. Bulk phase
corresponds to a triclinic unit cell which contains two C$_{22}H_{14}$
formulae with a set of parameters: $a=7.90$ {\AA}, $b=6.06$ {\AA},
$c=16.01$ {\AA}, $\alpha=101.9^\circ$, $\beta=112.6^\circ$, and
$\gamma=85.8^\circ$~\cite{Campbell}. Our calculated values of $a=7.90$ {\AA},
$b=6.06$ {\AA}, $c=16.01$ {\AA}, $\alpha=102.0^\circ$, $\beta=112.6^\circ$,
and $\gamma=85.5^\circ$ shows a very good agrement with the experimental
results of Campbell \textit{et al.} For the thin film phase, we obtained
an orthorhombic unit cell with $a=7.42$ {\AA}, $b=5.87$ {\AA}, and $c=16.21$
{\AA}. These parameters agree with Parisse \textit{et al.}'s~\cite{Parisse}
theoretical results of $a=7.60$ {\AA}, $b=5.90$ {\AA}, and $c=15.43$ {\AA}
except for the last one, corresponding to the longitudinal size of the unit cell.
We calculated the length of an isolated molecule to be 14.5~{\AA}.
Therefore, the disagreement, in this phase, stems from the Pn--Pn separation
in the multilayers.

In order to compare the predictions of different DFT functionals with the
experimental results for key structural and electronic parameters like the
lattice constants and binding energies, we repeated our calculations using
PW91~\cite{pw91} parametrization within both generalized gradient and local
density approximations (LDA). We calculated the lattice parameter of the silver
\textit{ccp} bulk structure in $Fm\bar3m$ symmetry group to be 4.00, 4.15, and
4.17 {\AA} using LDA-PW91, GGA-PW91, and GGA-PBE functionals respectively. These
compare well with the experimental~\cite{Moruzzi} value of 4.09 {\AA}, slightly
better than the previous theoretical~\cite{Li} result of 4.20 {\AA}. The Ag(111)
surface has been modeled in a four-layer slab geometry separated from their
periodic images by $\sim$15~{\AA} of a vacuum space and 3$\times$6$\times$1
grid for the cases of 1ML to 4ML deposition, whereas in the cases of isolated Pn
adsorption a larger cell is needed and therefore a three-layer-slab geometry for
the Ag(111) surface and 1$\times$2$\times$1 grid was used. For this metallic system,
the number of layers has been found to be sufficiently large to represent the
Ag(111) surface structure such that the geometry optimization calculations do
not disturb the subsurface layer atoms from their bulk lattice positions.

\begin{figure}
\epsfig{file=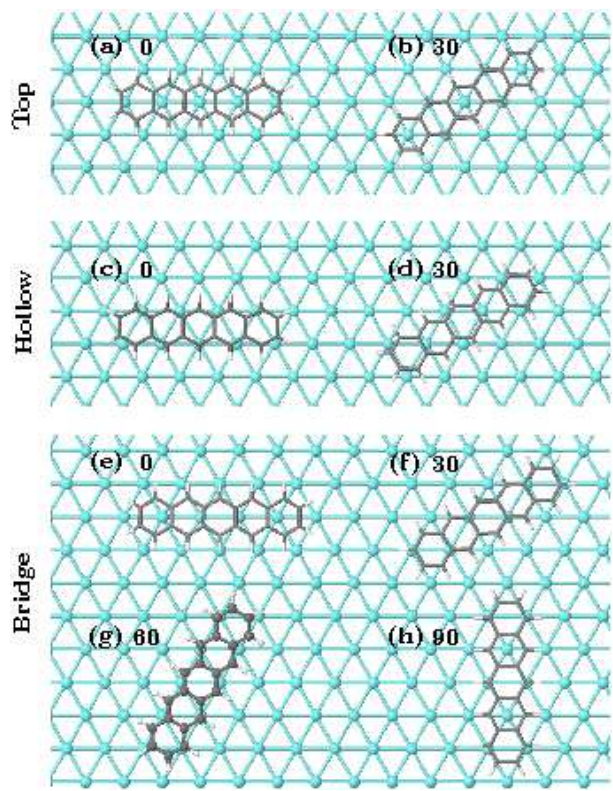,width=8cm}
\caption{Single isolated pentacene on different adsorption sites of Ag(111) surface.
Planar Pn adsorption with the central carbon ring on top of a silver atom aligning
parallel to one of the lattice directions is abbreviated as ``Top-0'' (a). Top-30 in
(b) refers to the adsorption at the top site with Pn major axis making an angle of
30$^\circ$ with any of the silver rows. Hollow-0 (c) and Hollow-30 (d) follow the
same molecular alignments as the top cases, but centered on a triangle whose corners
defined by Ag atoms, i.e. at the hollow site. Bridge-0 (e), Bridge-30 (f), Bridge-60
(g), and Bridge-90 (h), describe the cases where central ring of Pn lies over an
Ag--Ag bond making the referred angles with any of the lattice lines. The minimum
energy geometry, Bridge-60 (g), is depicted in ball-and-stick fashion while the
others are all shown in sticks only.
\label{fig1}}
\end{figure}

\section{Results and Discussion}

\subsection{Single isolated pentacene on Ag(111)}

In order to study the formation of ordered Pn layers on Ag(111) we first
considered a single Pn on the surface. Isolation of the molecules was
achieved by using a 8$\times$5 silver surface unit cell which sets 9.5 {\AA}
tip-to-tip and 7.9 {\AA} side-to-side separations between the periodic images.
We determined the minimum energy Pn/Ag(111) geometry by investigating all
possible adsorption sites with a number of orientations, compatible with
the lattice symmetry, at each site as shown in \ref{fig1}. (Labeling
conventions are described in figure caption.) In addition to these planar
cases we also investigated the possibility of standing-up adsorption
configurations which appeared to be around 0.15 eV less favorable than
the planar ones. In geometry optimization calculations Pn develops a weak
interaction with Ag(111) wherever it is initially placed on the surface.
In fact, as presented in \ref{table1}, the comparison of the total energies
of these adsorption cases show differences which are no greater than 36 meV
from each other. The flatness of the potential energy surface (PES) is
indicated by the existence of such small barriers which might make Pn
diffusion over the surface possible in agreement with the experimental
observations that the contact layer Pn molecules are mobile at the
Pn/Ag(111) interface~\cite{Eremtchenko,Dougherty}. Similarly, during image
acquisition STM tip has been observed to drag Pn molecules which are
physisorbed on Au(111)~\cite{Soe}.

\begin{table}
  \caption{Calculated values for geometrical and electronic structure of
  Pn/Ag(111) systems shown in \ref{fig1}. The lateral height of isolated
  Pn molecule from the Ag(111) surface $d_z$ in \AA, the binding energy
  $E_b$ and the relative total energy $E_T$ in eV.}
  \label{table1}
  \begin{tabular}{lccc} \hline\hline
    & $d_z$ & $E_b$ & $E_T$ \\[0.4mm] \hline &&& \\[-4mm]
    Top-0     & 3.90 & $-0.125$ & 0.030 \\
    Top-30    & 3.89 & $-0.119$ & 0.036 \\
    Hollow-0  & 3.88 & $-0.147$ & 0.008 \\
    Hollow-30 & 3.87 & $-0.128$ & 0.027 \\
    Bridge-0  & 3.87 & $-0.124$ & 0.031 \\
    Bridge-30 & 3.88 & $-0.124$ & 0.031 \\
    Bridge-60 & 3.87 & $-0.155$ & 0.000 \\
    Bridge-90 & 3.88 & $-0.129$ & 0.026 \\
    \hline\hline
    \end{tabular}
\end{table}

The adsorption configurations (in \ref{fig1}) where the isolated
pentacene follows the lattice symmetry so that the molecular
charge density matches better with the surface charge density
of silver rows are energetically more preferable. As a result,
the total energy of the bridge-60 is smaller from that of
the hollow-0 by only 8 meV. This also indicates that the
flatness of the PES is relatively more pronounced along the
lattice directions.

Single isolated pentacene molecule finds its minimum energy
configuration at the bridge-60 position as depicted in
ball-and-stick form in \ref{fig1}g. For this adsorption
geometry, it is almost flat with a negligible bending at a
height of 3.87 {\AA} which gives a weak binding energy of
$-0.155$ eV. In geometry optimization calculations, for all
possible initial adsorption configurations, both GGA
 functionals predict a weak interaction between the Pn 
molecule and the Ag(111) surface where LDA overbinds. 
(see \ref{table2}). In particular, GGA-PBE predicts that an 
isolated pentacene with a tilt about 15 degrees off the 
surface plane is only 3 meV unfavorable than the lowest
energy flat geometry. This barrier is so small that the 
tilted pentacene does not relax back to planar geometry.

Although GGA functionals result in a weak pentacene--silver
interaction, the degree of this weakness is overestimated.
Since, pure DFT results depend on the choice of the
exchange--correlation functional, a hybrid-DFT with corrected
exchange with dispersive interaction energy would give an improved
description of the binding characteristics of such a weakly bound
system.~\cite{Murdachaew}

\begin{table}
\caption{Calculated values for electronic and geometrical structure
of Ag and Pn/Ag(111) systems in different exchange--correlation
functionals. Lattice parameter of Ag(111) slab $a_{\rm Ag}$ in {\AA},
lateral heights $d_z$ (\AA) and the binding energies $E_b$ (eV) of
isolated and 1 ML Pn on the Ag(111) surface.\label{table2}}
\begin{tabular}{lccccc}
\cline{2-6}  &&&&& \\[-4mm]
\multicolumn{1}{c}{} & clean slab & \multicolumn{2}{c}{isolated Pn} & \multicolumn{2}{c}{1 ML Pn} \\
\cline{2-6} &&&&& \\[-4mm]
\multicolumn{1}{c}{} & $a_{\rm Ag}$ & $d_z$ & $E_b$ & $d_z$ & $E_b$ \\[0.4mm]
\hline &&&&& \\[-4mm]
LDA-PW91 & 4.000 & 2.46  & $-$1.925  & 2.48  & $-$1.753   \\
GGA-PW91 & 4.145 & 3.69  & $-$0.234  & 3.94  & $-$0.093   \\
GGA-PBE  & 4.174 & 3.87  & $-$0.155  & 4.12  & $-$0.078   \\
\hline
\end{tabular}
\end{table}

\subsection{Full monolayer}

\begin{figure}[htb]
\epsfig{file=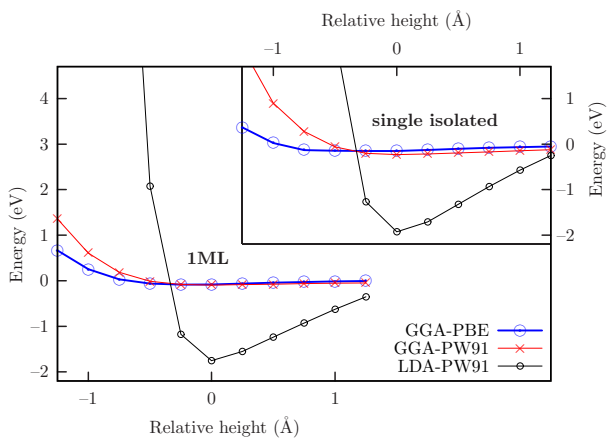,angle=0,width=8cm}
\caption{Binding energy versus pentacene height on Ag(111)
relative to the minimum energy position (bridge-60), calculated with
different exchange--correlation functionals and approximations
both for a single isolated Pn and for 1ML Pn coverage.\label{fig2}}
\end{figure}

Full monolayer has been derived from the previously optimized
bridge-60 configuration using an experimentally observed 6$\times$3
silver surface unit cell~\cite{Casalis,Danisman,Dougherty}.
The relaxed geometry of this contact layer is shown in
\ref{fig3}a--c. The molecules follow the surface symmetry
and are aligned parallel to the silver rows. The distance between
Pn and Ag(111) at the interface varies with varying intermolecular
interaction strengths. For instance, in the case of GGA-PBE, an
isolated Pn stays 3.9 {\AA} above the surface while this value
extends to 4.1 {\AA} in the case of 1 ML coverage as presented
in \ref{table2}. Corresponding binding energy at 1ML is calculated
to be 0.08 eV for GGA-PBE. This shows a weak binding similar to the
experimental observations.~\cite{Casalis,Danisman,Eremtchenko,Dougherty}
GGA-PW91 gives a slightly better description of both the Ag substrate
and Pn contact layer geometries relative to GGA-PBE results. We also
calculated the supercell total energies as a function of Pn--Ag(111)
distance as shown in \ref{fig2} for isolated and full monolayer cases.
LDA incorrectly gives strong binding for this system since the charge
densities are well localized around atoms leaving nearly empty
interatomic regions.

Our DFT calculations show that a small tilt angle of the contact 
layer will not yield a significant increase in the total energy 
with respect to flat geometry. This is due to the overestimated 
weakness of pentacene--silver interaction. Based on these DFT 
results we can not reject the possibility of an average tilt at 
the 1ML as well as the isolated single pentacene case depending 
on the experimental conditions. However, our calculations do not 
suggest a strong binding between the pentacene layer and the 
surface. Therefore, our calculations indicate a flat 1 ML 
physisorption rather than a tilted chemisorbed Pn layer which 
was concluded by K\"{a}fer \textit{et al.}~\cite{Kafer2} based 
on their NEXAFS and thermal desorption signatures.

\subsection{Multilayers}

\begin{figure}[htb]
\epsfig{file=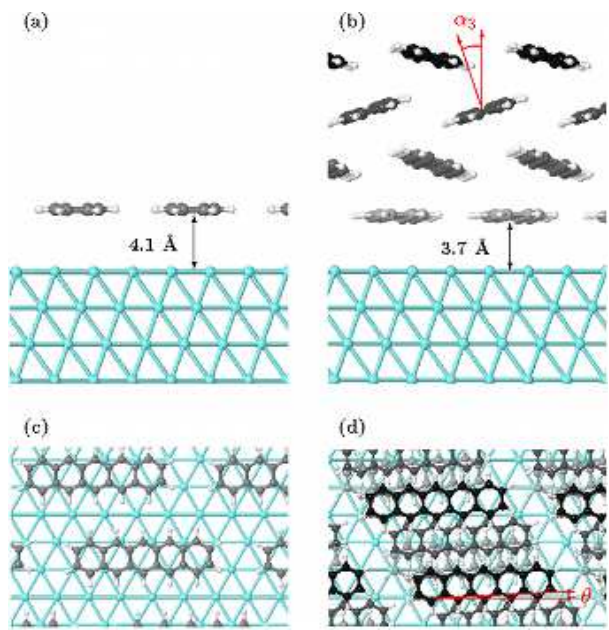,width=8cm}
\caption{Side views (a, b) and top views (c, d) of (1ML, 4ML) pentacene,
respectively, adsorbed on the Ag(111) surface. No perspective depth
is used in the production of the figures, hence the actual tilts are
seen in (b) and (d) for the 4ML case.\label{fig3}}
\end{figure}

We considered two different initial geometries for the second
pentacene layer on the already optimized 1ML Pn/Ag(111) structure.
For the first case, the second layer molecules are flat on the
first monolayer where molecular axes follow the surface symmetry.
In addition, a second layer pentacene stays above in between the two
molecules underneath. The second case initial structure is the same
as the first one except the molecular planes of only the second layer
pentacenes are tilted around their long axes as observed in experimental
studies.\cite{Kafer2,Casalis,Danisman,Eremtchenko,Dougherty}Geometry
optimization calculations resulted in a very small difference
of 0.046 eV in the total energies in favor of the latter case in
which second layer molecules slightly misaligned from the surface
lattice direction by 6.4$^\circ$ in addition to the molecular plane
tilting of 18$^\circ$ as presented in \ref{table3}. The smallness of
the energy difference between the two cases can be addressed to the
energy difference between the different phases of pentacene. For
instance, we calculated the difference in the total energies between
the bulk and the thin film phases of pentacene to be 0.12 eV for a
cell having two molecular formulae units.

\begin{figure}[htb]
\epsfig{file=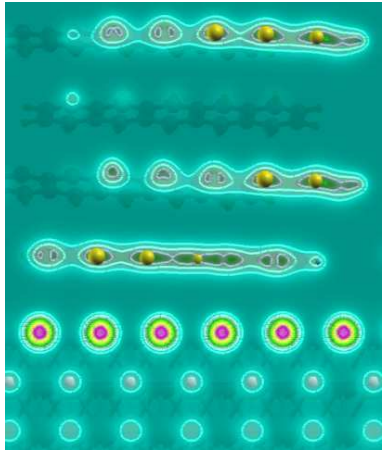,width=5cm}
\caption{The charge density contour plot of 4 ML pentacene--Ag(111) interface on a
plane normal to the surface and cut through a silver row.\label{fig4}}
\end{figure}

For the 2ML structure, the tilted layer (the second one) relaxes to
a height of 3.6 {\AA} above the contact layer which is separated
from the silver surface by 3.8 {\AA}. Resulting height of the first
tilted pentacene layer from silver surface becomes 7.4 {\AA}. In
the case of 3ML and 4ML structures this height converges to 7.2 {\AA}
which is slightly lower than the experimental value of 7.8 {\AA}
reported by Danisman \textit{et al.}.

\begin{table}[htb]
\caption{Calculated values for geometrical structure of Pn/Ag(111) systems
where $d_z$ is the distance between first-layer pentacene and Ag(111) surface,
$d_{n-m}$ is the distance between $n$th and $m$th Pn layers (all in \AA).
$\theta$ and $\alpha_n$ are the tilt angles of the $n$th layer molecules about the (111)-axis
and about their major axes, respectively.\label{table3}}
\begin{tabular}{lcccccccc}
\hline &&&&& \\[-4mm]
& $d_z$ & $d_{1-2}$ & $d_{2-3}$ & $d_{3-4}$ & $\theta$ & $\alpha_2$ & $\alpha_3$ & $\alpha_4$ \\[0.4mm]
\hline &&&&& \\[-4mm]
isolated & 3.9 &  -- &  -- &  -- & 0.0 & -- & -- & -- \\
1ML      & 4.1 &  -- &  -- &  -- & 0.0 & -- & -- & -- \\
2ML      & 3.8 & 3.6 &  -- &  -- & 6.4 & 18 & -- & -- \\
3ML      & 3.7 & 3.5 & 3.6 &  -- & 6.0 & 25 & $-$17 & -- \\
4ML      & 3.7 & 3.5 & 3.5 & 3.6 & 4.0 & 20 & $-$23 & 15 \\
\hline
\end{tabular}
\end{table}

The flat and tilted pentacene configurations above the first
layer has also been considered for the 3ML and 4ML initial
structures. The difference in the total energies has been
calculated to be 0.127 eV and 0.125 eV in favor of the tilted
molecules on the first layer for 3ML and 4ML cases, respectively.
Therefore, bulk-like pentacene formation on Ag(111) surface
above the contact layer is more preferable than flat lying
multilayers.

In the case of multilayers, the separation of the top layer
from the layer underneath is 3.6 {\AA} and all the inner
interlayer distances become 3.5 {\AA} while the height of the
contact layer converges to a value of 3.7 {\AA} after the
third ML. In addition, our calculations show a decrease in
the misalignment of pentacenes above the contact layer as new
layers deposited on Ag(111) surface up to 4ML. All multilayer
geometries can be seen through \ref{fig3}b--d, which only
show views of the 4ML structure, since the interlayer distances
and angles do not change significantly as new layers added.
We also present the corresponding electronic density contour 
plot of pentacene--Ag(111) interface at 4ML in \ref{fig4} which 
indicates charge localization around pentacene molecules and 
in the substrate suggesting weak binding of the contact layer.

Consequently, the thin film formation starting from the second
ML indicates that the intermolecular interactions are relatively
stronger than the pentacene--silver interaction. Substantiatively,
the misalignment of pentacenes above the contact layer can also
be attributed to the weakness of Pn--Ag(111) interaction. In fact,
this observation is in parallel with the experimental observations
of Danisman et al.~\cite{Danisman2} whereupon desorption of the
pentacene multilayers on a stepped Ag(111) surface a new monolayer
phase was observed. In addition, tilted second layer structure was
also reported by both Eremthcenko et al.\cite{Eremtchenko} and
K\"{a}fer et al.~\cite{Kafer2} Furthermore, molecular rearrangements
involving such topological phase transformations from flat to
buckled pentacene multilayers mimicking thin film formation on
Ag(111) can be expected to give the same order of energy differences
as that of thin film and the bulk pentacene phases. One final point
to be stressed here is that, although (i) our results may be helpful
for the comparison of stability of flat and tilted multilayers, and
(ii) the more stable multilayer configurations we found resemble the
experimentally observed pentacene phases\cite{Kafer2,Casalis,Danisman,Nabok,Campbell}
(i.e., the tilt angles are very close to that of pentacene bulk and
thin film phases), a direct comparison of our results with the
experimentally observed structures may not be very meaningful. This
is because the multilayer and monolayer in-plane unit-cell dimensions
which are actually different had to be chosen the same due to
computational restrictions.

\begin{figure}[htb]
\epsfig{file=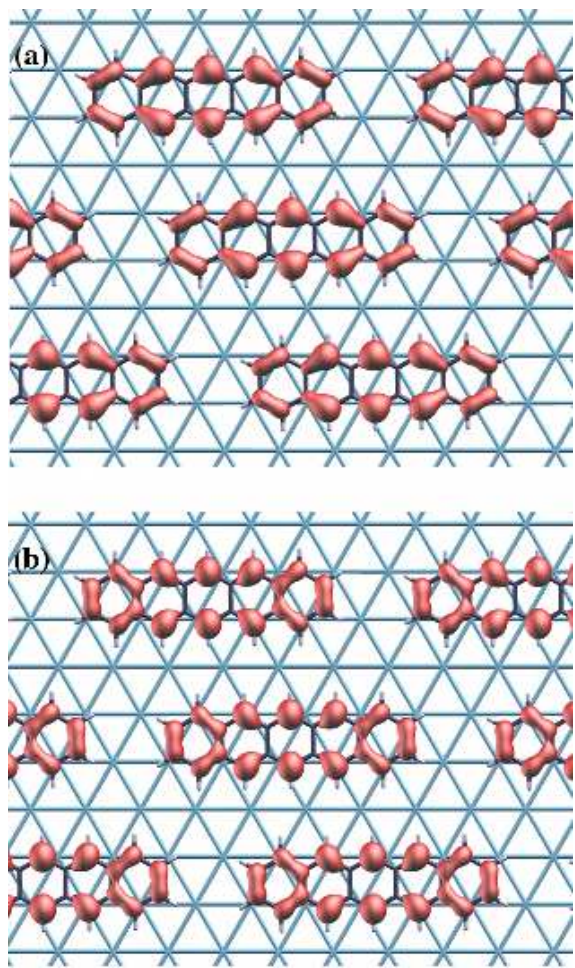,width=7.5cm}
\caption{Calculated STM image of 1ML Pn on Ag(111) (a) for the
occupied states around $-0.9$ eV and (b)for the unoccupied states
around 1.3 eV vicinity of the Fermi energy. \label{fig5}}
\end{figure}

\begin{figure}[htb]
\epsfig{file=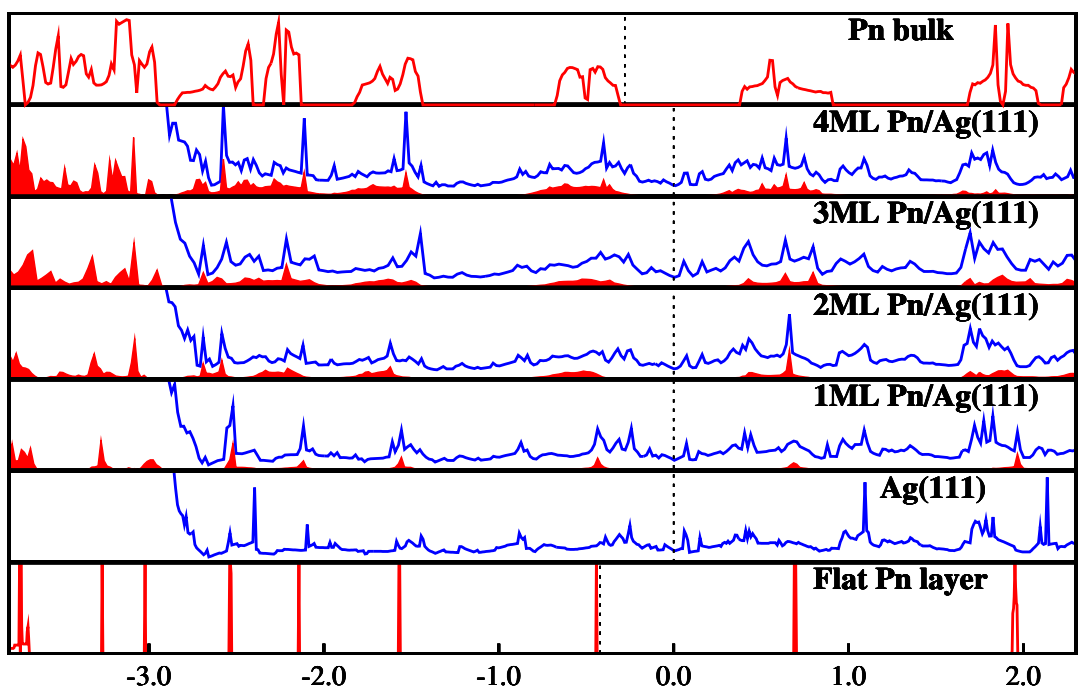,angle=-90,width=15cm}
\caption{(color online) Calculated PDOS for Pn/Ag(111) structures.
The abscissa is the energy, in eV, relative to the Fermi level for
the clean Ag(111) surface. Pentacene contributions are indicated
by gray (red) while total DOS is in dark gray (blue). \label{fig6}}
\end{figure}

In order to investigate the coupling of the frontier molecular
orbitals of pentacene to the silver substrate states, we obtained
the STM pictures by using Tersoff--Hamann approximation~\cite{Tersoff}.
The calculated STM images for the applied voltages of $-0.9$ V
and $1.3$ V in \ref{fig5} resemble to the HOMO and LUMO charge
densities of an isolated Pn similar to Lee \textit{et al.}'s
result~\cite{Lee1}. Our results also agree well with the recent
differential conductance images obtained with low-temperature STM
experiments for seemingly similar physisorption system of
pentacene/Au(111)~\cite{Soe}. These STM images in \ref{fig5} are
consistent and are also apparent from the PDOS of 1ML Pn/Ag(111)
presented in \ref{fig6}. The first PDOS peak of the Pn layer about
0.5 eV below the Fermi energy comes from the HOMO's of the molecules.
The sharpness of this peak substantiates that the frontier orbitals
of the Pn molecules mixes very weakly with the $5s$ states of the
surface Ag atoms. Hence, STM calculation covering an energy range
of 0.9 eV below the Fermi level shows HOMO charge density for Pn
layer over the slab as shown in \ref{fig5} where the small Ag $5s$
contribution was suppressed for visual convenience. Similarly, the
first peak due to Pn layer about 0.7 eV above the Fermi energy
corresponds to the LUMO's of the molecules which are also weakly
mixing with the valence bands of Ag(111). At this point, in order
to comment on the reliability of our method, we compare the experimental
and theoretical results of Pn--Cu(100). Ferretti et al.~\cite{Ferretti},
on this system, have found a significant broadening of pentacene
HOMO--LUMO bands and mixing with the substrate levels using the same
computational procedure. This, in combination with the experimental
results, were interpreted as an interaction close to chemisorption.
In Pn--Ag(111) case, however, it is clear from our results that the
picture is more close to physisorption.

The bottom panel of \ref{fig6} shows the calculated DOS for the
flat pentacene layer which is obtained by removing the silver
substrate from 1ML/Ag(111) (\ref{fig3}a--c) structure. The
sharp peaks, having less structure, rather look like an energy
level diagram due to very low overlap between molecular orbitals
through tip to tip pentacene contacts over the layer. Metallic
nature of the bare Ag(111) surface is also presented in the
succeeding panel of \ref{fig6} where Fermi level is set as
the origin of the energy axis. The DOS structure corresponding
to full pentacene monolayer on Ag(111) is shown in the third
panel from the bottom in \ref{fig6}. DOS peaks stemming from
pentacene show no shift in energy with respect to those of the
flat Pn layer in the absence of the metal substrate. In addition,
the broadening of each peak is localized over a small number
of silver states indicating weak semiconductor--metal coupling.
As a result of this weak interaction, the contact layer shows
bulk-like HOMO--LUMO contributions to the total DOS around the
Fermi energy. Therefore, STM experiments capture these frontier
molecular orbitals.

As new pentacene layers deposited on the first full monolayer
the corresponding PDOS contribution starts to form localized
satellite structures at around flat Pn layer peak positions.
Their broadening is larger than the broadening in PDOS peaks
obtained for 1ML/Ag(111). This indicates that the interlayer
molecular orbital overlap is relatively stronger than the
coupling between the contact layer and the metal substrate.
Moreover, these PDOS satellites match perfectly with the
bulk pentacene DOS which is presented in the top panel
of \ref{fig6}. Therefore, energetically preferable thin film
pentacene phase on Ag(111) up to 4ML possesses bulk-like DOS
properties.

Our DOS calculations show that pentacene has no electronic
contribution at the Fermi energy. Evidently, highly ordered
pentacene multilayers on Ag(111), considered in this study,
does not exhibit band transport. In addition, these multilayers
occur in bulk-like phase where the overlap of the molecular
orbitals between nearest neighbor pentacenes yield large
$\pi$-conjugation length along the molecular axis. Therefore,
our results suggest a hopping mechanism between the localized
states for the carrier transport.

\section{Conclusion}

In conclusion, we have investigated the geometric and electronic
structure of pentacene on Ag(111) surface up to 4 ML coverage at
the DFT level where GGA functionals perform better. At the most
stable configuration a single isolated pentacene lies flat at
3.9 {\AA} above Ag(111) surface on the so called Bridge-60
position with a weak binding energy of $-$0.155 eV. For the full
monolayer coverage, molecules align perfectly with the silver
rows on the surface while the ML height extends slightly to
4.1 {\AA} due to intermolecular interactions. Calculated binding
energies as well as STM and PDOS structures indicate weak
pentacene--substrate coupling.

Pentacenes above the contact layer favor the thin film multilayer
structure over the planar configuration with a slight energy
difference which can be addressed to the small energy barriers
between different phases of pentacene. Moreover, the slight
misalignment of pentacene molecules above the first layer from
the surface silver rows indicate that a bulk-like thin film phase
starting from the second layer is adsorbed on the Ag(111) surface
through a contact layer at the interface.

\end{document}